\documentclass[aps,twocolumn,prabib,amsmath,amssymb,superscriptaddress,notitlepage]{revtex4-1}
\usepackage{graphics}
\usepackage{bm}
\usepackage{dcolumn}
\usepackage{natbib}
\usepackage{epstopdf}
\usepackage{float}
\usepackage{graphicx}
\usepackage{epsfig}
\usepackage[pdfstartview=FitH]{hyperref}
\usepackage{color}
\usepackage{appendix}
\usepackage{ulem}

\newcommand{\rr}{{\bf r}}

\begin{document}

\title{Topological characterization of hierarchical fractional quantum Hall effects in topological flat bands with SU($N$) symmetry}
\author{Tian-Sheng Zeng}
\affiliation{School of Science, Westlake University, Hangzhou 310024, China}
\affiliation{Institute of Natural Sciences, Westlake Institute for Advanced Study, Hangzhou 310024, China}
\author{D. N. Sheng}
\affiliation{Department of Physics and Astronomy, California State University, Northridge, California 91330, USA}
\author{W. Zhu}
\affiliation{School of Science, Westlake University, Hangzhou 310024, China}
\affiliation{Institute of Natural Sciences, Westlake Institute for Advanced Study, Hangzhou 310024, China}
\date{\today}
\begin{abstract}
We study the many-body ground states of SU($N$) symmetric hardcore bosons on the topological flat-band model by using controlled numerical calculations.
By introducing strong intracomponent and intercomponent interactions,
we demonstrate that a hierarchy of bosonic SU($N$) fractional quantum Hall (FQH) states
emerges at fractional filling factors $\nu=N/(MN+1)$ (odd $M=3$).
In order to characterize this series of FQH states, we figure
the effective $\mathbf{K}$ matrix from the inverse of the Chern number matrix.
The topological characterization of the $\mathbf{K}$ matrix also reveals
quantized drag Hall responses and fractional charge pumping that could be detected in future experiments.
In addition, we address the general one-to-one correspondence to the spinless FQH states
in topological flat bands with Chern number $C=N$ at fillings $\widetilde{\nu}=1/(MC+1)$.
\end{abstract}
\maketitle

\section{introduction}
The topological Chern number is a topological invariant
that classifies the ground state of the quantum Hall systems~\cite{Thouless1982}.
Importantly, the nonzero Chern number indeed leads to experimentally measurable quantum phenomena with a topological origin,
e.g. quantized charge pumping~\cite{Thouless1983},
where the amount of pumped charge during an adiabatic cycle in periodic parameter space is determined by the topological Chern number.
This novel relationship provides a direct and flexible characterization of quantum Hall physics,
%beyond electronic systems,
which has been realized in advanced cold atom experiments owing to the
unprecedented level of control of superlattice systems.
The examples include a two-dimensional quantum Hall effect with linear Hall response~\cite{Yoshiro2016,Bloch2016,Jotzu2014},
four-dimensional quantum Hall effect with non-linear Hall response~\cite{Bloch2018},
and spin pumping of ultracold bosonic atoms in an optical superlattice~\cite{Schweizer2016}.

So far most studies of topological pumping have focused on the single component experiments.
In this case, charge transfer simply relates to the total Hall conductance of the system, without any
information about the internal topological structure of the system.
Multicomponent systems, however, necessitate a generalization of the fractional quantum Hall (FQH) theory that takes into account the mutual gauge fields between different components.
In particular, instead of a single quantum number, the topological information is encoded in an integer valued symmetric $\mathbf{K}$ matrix, which classifies the topological order at different particle fillings for Abelian multicomponent systems within the framework of the Chern-Simons theory~\cite{Wen1992a,Wen1992b,Blok1990,Blok1991}.
For example, Halperin's two-component $(mmn)$ wave function is described by the $\mathbf{K}=\begin{pmatrix}
m & n\\
n & m\\
\end{pmatrix}$ matrix~\cite{Halperin1983}, where the diagonal and off-diagonal terms describe
intraspecies and interspecies topological responses, respectively.

In condensed matter, the multicomponent quantum Hall effects have been realized in many different systems
including monolayer graphene, where the valley and spin degrees of freedom represent different components with approximate SU(4) symmetry~\cite{Bolotin2009,Du2009,Dean2011,Young2012,Feldman2012,Feldman2013,Kim2019}.
In addition, the $N$-component FQH states can be mapped into SU($N$) spin liquids and realized in quantum magnetism \cite{Laughlin1987,Tu2014,Bondesan2014,Tu2017}.
In such systems, the total Hall conductance is the commonly measured quantity containing topological information.
The Hall conductance, while providing a quantitative measure of total Chern number of a many-body ground state,
contains little information about how different components are entangled with each other.
Therefore, to gain a better understanding of the structure of multicomponent FQH states
~\cite{Balatsky1991,Lopez1995,Read1997,Lopez2001,Toke2007,Goerbig2007,Gail2008,Modak2011,Furukawa2012,YHWu2013,Grass2014,Toke2012,Abanin2013,Sodemann2014,Jolicoeur2014,Sterdyniak2015b,Reijnders2002,Reijnders2004},
the topological characterization based on a $\mathbf{K}$ matrix is highly desired.
Moreover, experimental realization of Chern insulators in cold atoms~\cite{Jotzu2014}
and bilayer graphene heterostructure~\cite{Spanton2017},
would open up an avenue for exploration of the multicomponent quantum Hall effects
~\cite{Sun2011,Neupert2011,Sheng2011,Tang2011,Wang2011,Regnault2011}.
Accordingly, numerical characterization of the topological nature of two-component Halperin (221) and (331) states from
the $\mathbf{K}$ matrix is derived from the inverse of the Chern number matrix in Ref.~\cite{Zeng2017}.
This topological characterization can be generalized to multicomponent SU$(N)$
FQH effects for bosons at $\nu=N/(N+1)$ and fermions at $\nu=N/(2N+1)$.
In addition, it has been numerically verified~\cite{Zeng2018}, that there is a close relationship between
multicomponent FQH states and color-entangled states at partial fillings $\widetilde{\nu}=1/(MC+1)$ ($M=1$
for hardcore bosons and $M=2$ for spinless fermions)~\cite{LBFL2012,Wang2012r,Yang2012,Sterdyniak2013,Wang2013,Moller2015,YLWu2013,YLWu2014,YHWu2015,Behrmann2016}
in lattice models forming bands with higher Chern number $N=C>1$. With this progress at hand, it is natural to ask if
the diagnosis of $\mathbf{K}$ matrix could identify some more FQH states, which motivates us to investigate FQH states at different filling series.

In this paper, we focus on the FQH physics for $N$-component hardcore bosons with SU$(N)$-invariant interactions
in a concrete topological lattice model at a filling factor $\nu=N/(3N+1)$.
To our best knowledge, the FQH effect at this filling series has not been numerically addressed before.
Through density matrix renormalization group (DMRG) and exact diagonalization (ED) calculations,
we show that a class of incompressible FQH states emerges at $\nu=N/(3N+1)$ under the interplay of interaction and band topology.
Topological properties of these states are characterized by the $\mathbf{K}$ matrix~\cite{Blok1990}.
Furthermore, an explicit calculation for systems with similar geometry as experiments
reveals the fractional quantization of the drag charge transfer,
which can be interpreted as the prime measurable physical consequence of the topological nature of multicomponent FQH effects.

This paper is organized as follows. In Sec.~\ref{model}, we introduce the multicomponent interacting Hamiltonian of hardcore bosons loaded on $\pi$-flux topological checkerboard lattice, and describe the general physical scheme to understand the physics of the $\mathbf{K}$ matrix of multicomponent SU$(N)$ FQH states from the inverse of the Chern number matrix. In Sec.~\ref{kkmatrix}, under strong SU$(N)$ symmetric interactions, we numerically demonstrate the emergence of a hierarchy of FQH effects at fillings $\nu=N/(3N+1)$ for hardcore bosons,
based on topological information of the $\mathbf{K}$ matrix,
including (1) fractionally quantized topological invariants related to Hall conductances, and
(2) a nearly degenerate ground state manifold.
In Sec.~\ref{fci}, we discuss the close relationship between these SU$(N)$ symmetric $N$-component FQH states
and the physics in topological flat bands with Chern number $N$ at fillings $\widetilde{\nu}=1/(MC+1)$. In Sec.~\ref{drag}, we discuss the fractional charge and spin pumping due to the quantized drag Hall conductance. Finally, in Sec.~\ref{summary}, we summarize our results and discuss the prospect of investigating nontrivial SU$(N)$ symmetric $N$-component FQH states.

\section{Models and Methods}\label{model}
We consider a system with $N$-component hardcore bosons with SU$(N)$-invariant interactions on the topological $\pi$-flux checkerboard lattice:
\begin{align}
  \quad\quad\quad&H=\sum_{\sigma}H_{CB}^{\sigma}+V_{int},\label{cbl}\\
  V_{int}=U&\sum_{\sigma\neq\sigma'}\sum_{\rr}n_{\rr,\sigma}n_{\rr,\sigma'}+V\sum_{\sigma,\sigma'}\sum_{\langle\rr,\rr'\rangle}n_{\rr',\sigma}n_{\rr,\sigma'},\label{interact}
\end{align}
where $H_{CB}^{\sigma}$ denotes the noninteracting Hamiltonian of the $\sigma$-th component $\sigma=1,2,\cdots,N$,
\begin{align}
  &H_{CB}^{\sigma}=-t\!\sum_{\langle\rr,\rr'\rangle}\!\big[b_{\rr',\sigma}^{\dag}b_{\rr,\sigma}\exp(i\phi_{\rr'\rr})+H.c.\big]\nonumber\\
  &-\!\sum_{\langle\langle\rr,\rr'\rangle\rangle}\!\!\! t_{\rr,\rr'}'b_{\rr',\sigma}^{\dag}b_{\rr,\sigma}
  -t''\!\sum_{\langle\langle\rr,\rr'\rangle\rangle}\!\!\!\! b_{\rr',\sigma}^{\dag}b_{\rr,\sigma}+H.c.,\label{cb}
\end{align}
Here $b_{\rr,\sigma}^{\dag}$ creates a boson of the $\sigma$-th component at site $\rr$, $n_{\rr,\sigma}=b_{\rr,\sigma}^{\dag}b_{\rr,\sigma}$ is the particle number operator
of the $\sigma$-th component at site $\rr$, and $\langle\ldots\rangle$,$\langle\langle\ldots\rangle\rangle$
and $\langle\langle\langle\ldots\rangle\rangle\rangle$ denote the first, second and third nearest-neighbor
pairs of sites.
In the flat-band limit, we use the parameters $t'=0.3t,t''=-0.2t,\phi=\pi/4$ for a checkerboard lattice~\cite{Zeng2015}.
Here we consider the on-site intercomponent and nearest-neighbor intracomponent interactions.
$U$ ($V$) is the strength of the SU$(N)$ symmetric onsite intercomponent interaction (the
nearest-neighbor intracomponent interaction).

In the ED calculations, the finite systems we study enclose
$N_x\times N_y$ unit cells (the total number of sites is $N_s=q N_x N_y$,
with $q$ the number of inequivalent sites within a unit cell).
We focus on the filling series $\nu=qN_e/N_s=N/(3N+1)$ (e.g., $\nu=2/7,3/10$) of the lowest Chern band,
where $N_e=\sum_{\sigma}N_{\sigma}$ is the total particle number.
%Within the translational symmetry,
%the energy eigenstates are labeled by the total momentum $K=(K_x,K_y)$ in the Brillouin zone.
For larger system sizes, we apply DMRG on a cylinder geometry with the open boundary condition in the $x$-direction
and the periodic boundary condition in the $y$-direction.
We keep the number of states up to $2400$ to ensure the maximal discarded truncation error is less than $10^{-5}$.

Generally speaking, the multicomponent FQH states
can be classified by a class of the integer-valued symmetric $\mathbf{K}$ matrix of the rank $N$ ~\cite{Wen1992a,Wen1992b,Blok1990,Blok1991}.
The diagnosis of the topological nature of the $\mathbf{K}$ matrix has been discussed in our previous works~\cite{Zeng2017,Zeng2018}.
Here we briefly summarize the main strategy.
For multicomponent FQH states at a given filling $\nu=N/(MN+1)$
(odd $M$ for hardcore bosons and even $M$ for fermions), the $\mathbf{K}$ matrix has the form
\begin{align}
  \mathbf{K}=\begin{pmatrix}
M+1 & M & \cdots & M\\
M & \ddots & \ddots & \vdots\\
\vdots & \ddots & \ddots & \vdots\\
M & \cdots &\cdots & M+1\\
\end{pmatrix},\label{kmatrix}
\end{align}
where the number of columns and rows is set by $N$.
Physically, this can be understood that the particles are attached to an $M$ number of flux quanta,
forming the composite fermions~\cite{Jain1989}. Bosons can be attached to odd $M$ fluxes to form a composite fermion, while fermions are attached to even $M$ fluxes to form a composite fermion. For odd $M=1,3,5,7\cdots$, the $\mathbf{K}$ matrix indicates a hierarchy of bosonic multicomponent FQH at sequential fillings $\nu=N/(N+1), \nu=N/(3N+1), \nu=N/(5N+1), \nu=N/(7N+1)$ and so on; for even $M=2,4,6,8\cdots$, the $\mathbf{K}$ matrix indicates a fermionic multicomponent FQH at $\nu=N/(2N+1), \nu=N/(4N+1), \nu=N/(6N+1), \nu=N/(8N+1)$ and so on.
The $\mathbf{K}$ matrix describes the precise nature of the internal topological nature of multicomponent FQH states.
For instance, from Eq.~\ref{kmatrix}, we can extract the determinant $\det\mathbf{K}=MN+1$, which characterizes the topological degeneracy of the ground state manifold.
Under a special linear group SL($N,\mathbb{Z}$) transformation,
Eq.~\ref{kmatrix} is related to the Cartan matrix of the Lie algebra SU$(N)$, indicating these FQH ground states host a SU$(N)_1$ Kac-Moody symmetry~\cite{Zee1991,Wen1992a}.
The $\mathbf{K}$ matrix is related to their multicomponent Hall conductance responses (denoted by the Chern number matrix $\mathbf{C}$ for a multicomponent system), through
\begin{align}
  \mathbf{C}=\mathbf{K}^{-1}=\begin{pmatrix}
C_{1,1} & C_{1,2} & \cdots & C_{1,N}\\
C_{2,1} & \ddots & \ddots & \vdots\\
\vdots & \ddots & \ddots & \vdots\\
C_{N,1} & \cdots &\cdots & C_{N,N}\\
\end{pmatrix}.\label{chern}
\end{align}
Here the diagonal part of matrix elements $C_{\sigma,\sigma}$ represents the intracomponent Hall conductance (in unit of conductance quanta $e^2/h$),
where the off-diagonal part $C_{\sigma,\sigma'}$ is related to the intercomponent drag Hall conductance between particles of the $\sigma$-th and
the $\sigma'$-th components.
For our SU$(N)$ symmetric systems, each component contributes the same charge amount as unit.
Thus the $N$-component charge vector can be taken as $\mathbf{q}^T=[1,1,\cdots,1]$, and the total charge Hall conductance is given by~\cite{Wen1995}
\begin{align}
  \sigma_H=\mathbf{q}^{T}\cdot\mathbf{K}^{-1}\cdot\mathbf{q}=\sum_{i,j}C_{i,j}=\nu.\nonumber
\end{align}

\begin{figure}[t]
  \includegraphics[height=2.75in,width=3.4in]{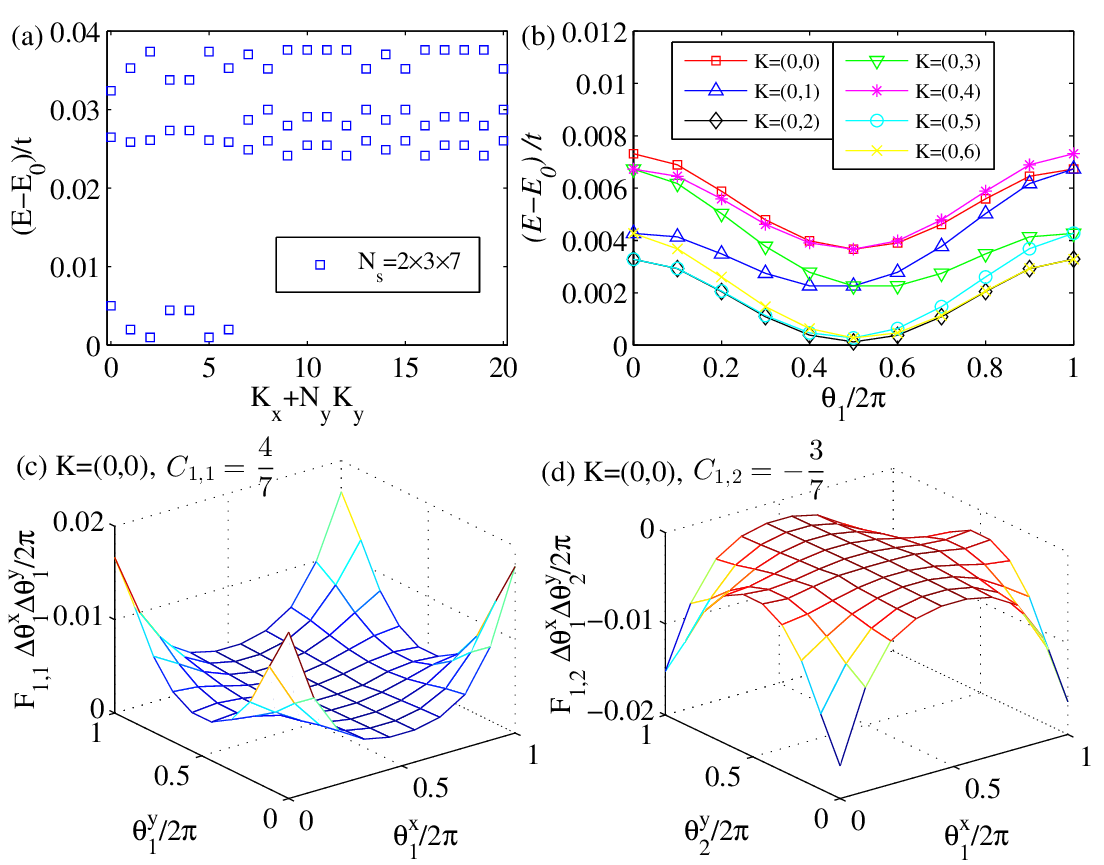}
  \caption{\label{energy}(Color online) Numerical ED results for two-component hardcore bosons with  $\nu=2/7,N_s=2\times3\times7,U=0,V=10t$ on the topological $\pi$-flux checkerboard lattice: (a) the low energy spectrum in each momentum sector; (b) the $y$ direction spectral flow of degenerate ground state energy levels under the insertion of flux quantum $\theta_{1}^{y}=\theta_1,\theta_{2}^y=0$. Berry curvatures $F_{\sigma,\sigma'}^{xy}\Delta\theta_{\sigma}^{x}\Delta\theta_{\sigma'}^{y}/2\pi$ of the $K=(0,0)$ ground state in the (c) $(\theta_{1}^{x},\theta_{1}^{y})$ and (d) $(\theta_{1}^{x},\theta_{2}^{y})$ parameter planes.}
\end{figure}

The interacting many-body Hall conductance can be obtained by by revealing the many-body counterpart of the Chern number~\cite{Niu1985}. Numerically, the Chern number matrix can be calculated using the twisted boundary scheme~\cite{Sheng2003,Sheng2006}.
Under twisted boundary conditions $\psi(\cdots,\rr_{\sigma}^{i}+N_{\alpha}{\hat e_{\alpha}},\cdots)=\psi(\cdots,\rr_{\sigma}^{i},\cdots)\exp(i\theta_{\sigma}^{\alpha})$
where $\theta_{\sigma}^{\alpha}$ is the twisted angle for particles of the $\sigma$-th component in the $\alpha$-direction.
In a two-parameter $(\theta_{\sigma}^{x},\theta_{\sigma'}^{y})$ plane,
we can define the many-body Chern number $C_{\sigma,\sigma'}$ of the ground state wavefunction $\psi$ through the integral of the Berry curvature $F_{\sigma,\sigma'}^{xy}$ ~\cite{Sheng2003,Sheng2006}
\begin{align}
  C_{\sigma,\sigma'}=\frac{1}{2\pi}\int d\theta_{\sigma}^{x}d\theta_{\sigma'}^{y}F_{\sigma,\sigma'}^{xy},\label{chern1}
\end{align}
and
\begin{align}
    F_{\sigma,\sigma'}^{xy}=\mathbf{Im}\left(\langle{\frac{\partial\psi}{\partial\theta_{\sigma}^x}}|{\frac{\partial\psi}{\partial\theta_{\sigma'}^y}}\rangle
-\langle{\frac{\partial\psi}{\partial\theta_{\sigma'}^y}}|{\frac{\partial\psi}{\partial\theta_{\sigma}^x}}\rangle\right).\nonumber
\end{align}
For $M=1,2$, the $\mathbf{K}$ matrices have been successfully applied to the diagnosis of SU$(N)$ FQH states at a series of fillings $\nu=N/(MN+1)$ ~\cite{Zeng2017,Zeng2018}.
For $M>2$, however, to our best knowledge there are no studies of
such states in microscopic systems, such as $M=3$, which is the focus of the present work.

\begin{figure}[t]
  \includegraphics[height=2.0in,width=3.2in]{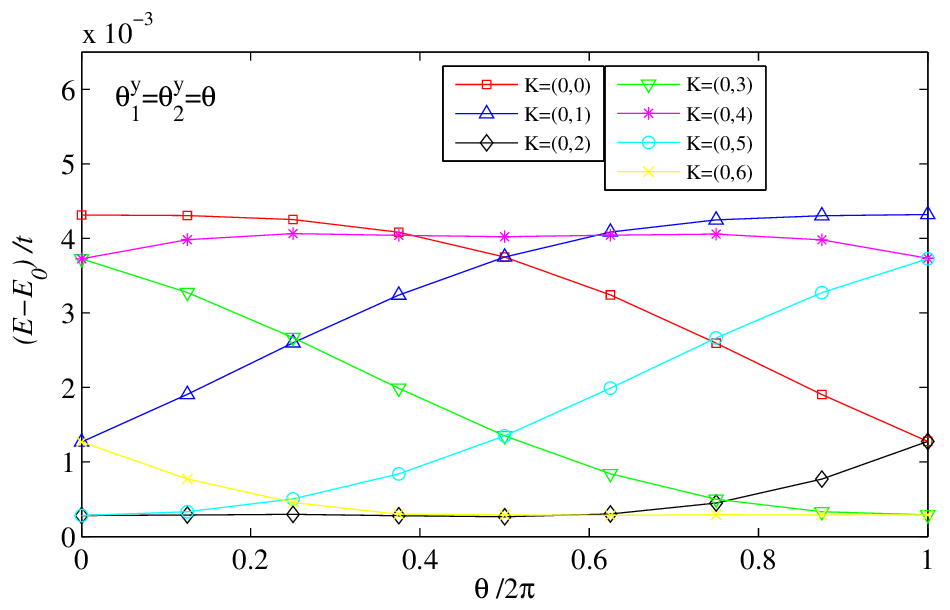}
  \caption{\label{flux2}(Color online) The $y$ direction spectral flow of degenerate ground state energy levels for two-component hardcore bosons with  $\nu=2/7,N_s=2\times3\times7,U=0,V=10t$ on the topological $\pi$-flux checkerboard lattice under the insertion of flux quantum $\theta_{1}^{y}=\theta_{2}^y=\theta$.}
\end{figure}

\section{SU$(N)$ FQH states}\label{kkmatrix}

We now discuss the numerical evidence of the emergence of multicomponent FQH states at a given filling $\nu=N/(MN+1)$ (e.g., $\nu=2/7$ for $M=3,N=2$) under strong SU$(N)$ symmetric interactions. Due to the difficulty of ED study for $N>2$, we provide the proof of SU$(N>2)$ FQH states from DMRG simulation of drag charge pumping in Sec.~\ref{drag}.
%Now we begin to discuss the numerical evidences of SU$(2)$ bosonic FQH states at $\nu=2/7$ for $M=3$.

First, in Figs.~\ref{energy}(a) and~\ref{energy}(b), we plot the low-energy spectrum of strong interacting hardcore bosons at $\nu=2/7,U=0,V=10t$.
The key finding in this calculation is that, there is a seven-fold quasidegenerate ground state manifold separating from higher-energy levels by a robust gap.
We have checked that this degeneracy manifold persists for $U\gg t$.
This seven-fold degeneracy is consistent with the prediction from the determinant of $\mathbf{K}$ matrix $\det\mathbf{K}=2M+1=7$.

Further, in Fig.~\ref{energy}(b), we plot the low-energy spectrum evolution under the insertion of flux quantum denoted by $\theta_{\sigma}^{\alpha}$.
The seven-fold ground state manifold evolves into each other without gap closing during each cycle.
Interestingly,
the energy spectrum evolves back into itself
after the insertion of seven flux quanta for both $\theta_{1}^{\alpha}=\theta_{2}^{\alpha}=\theta$,
and $\theta_{1}^{\alpha}=\theta,\theta_{2}^{\alpha}=0$ [as indicated in Figs.~\ref{energy}(b) and~\ref{flux2}, respectively],
indicating that elemental quasiparticles take a minimally fractionalized $1/7$-statistics of a physical hardcore boson.
This is an evidence that the $\nu=2/7$ state is an Abelian FQH state.

\begin{figure}[t]
  \includegraphics[height=1.51in,width=3.4in]{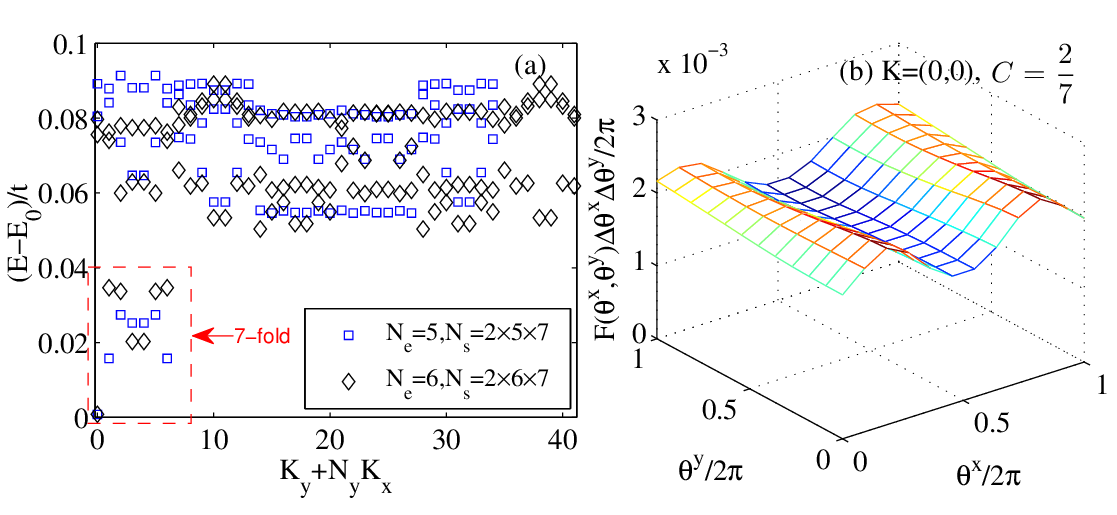}
  \caption{\label{energy2}(Color online) Numerical ED results for (a) the low-energy spectrum of hardcore bosons $\widetilde{\nu}=1/7,N_s=2\times N_e\times7$ at $U=10t,V=10t$ on the square lattice with Chern number $C=2$ and (b) the corresponding Berry curvatures of the $K=(0,0)$ ground state in the plane $(\theta^{x},\theta^{y})$.}
\end{figure}

Moreover, we study the Berry curvatures carried by the seven ground states obtained in ED calculations.
In numerics, we use discrete $m\times m$ meshes in the boundary phase space with $m\geq10$: $\Delta\theta^\alpha_\sigma=2\pi /m$.
As an example, the Berry curvatures $F_{1,1},F_{1,2}$ of the ground state at momentum sector $K=(0,0)$
are shown in Figs.~\ref{energy}(c) and~\ref{energy}(d), respectively.
Importantly, the sum of Berry curvatures can give rise to fractionally quantized Chern numbers $C_{i,j}$. Accordingly,
we obtain integer quantized invariants $\sum_{i=1}^{7} C_{1,1}^i=4$, and $\sum_{i=1}^{7} C_{1,2}^i=-3$ for these seven-fold degenerate ground states at momenta $K=(0,i)$.
All of the above imply a $2\times2$ $\mathbf{C}$ matrix,
\begin{align}
  \mathbf{C}=\begin{pmatrix}
C_{1,1} & C_{1,2}\\
C_{2,1} & C_{2,2}\\
\end{pmatrix}=\frac{1}{7}\begin{pmatrix}
4 & -3\\
-3 & 4\\
\end{pmatrix}.
\end{align}
Finally the $\mathbf{K}$ matrix can be obtained from the inverse of the $\mathbf{C}$ matrix,
namely $\mathbf{K}=\mathbf{C}^{-1}=\begin{pmatrix}
4 & 3\\
3 & 4\\
\end{pmatrix}$.
Therefore, from the above three aspects, we faithfully
establish the topological nature of $\nu=2/7$ FQH states as a lattice version of Halperin (443) states.

\section{Color-entangled FQH effects in topological bands with Chern number $N$}\label{fci}

\begin{figure}[t]
  \includegraphics[height=1.6in,width=3.4in]{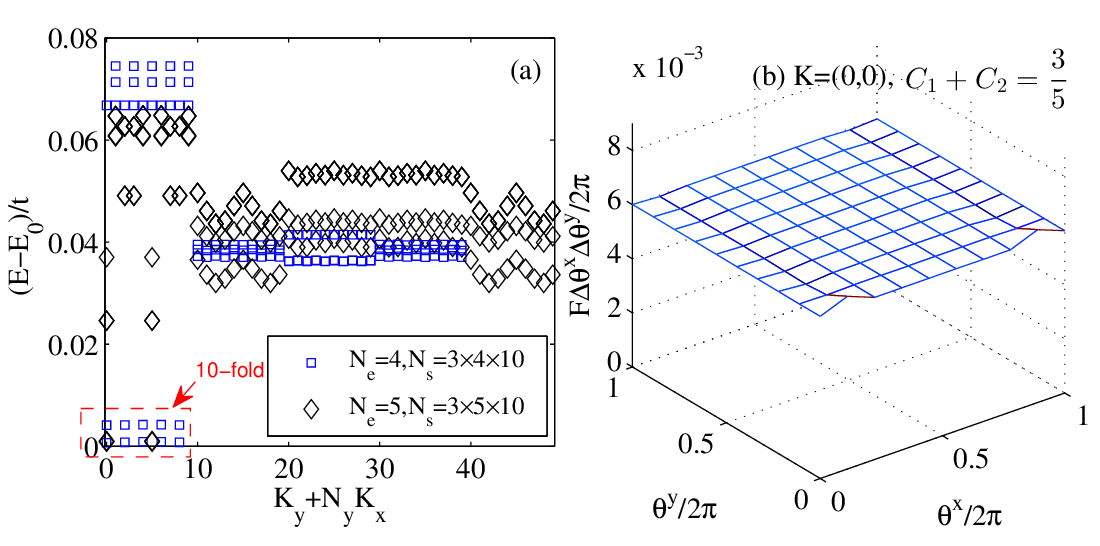}
  \caption{\label{energy3}(Color online) Numerical ED results for (a) the low-energy spectrum of hardcore bosons
  $\widetilde{\nu}=1/10,N_s=2\times N_e\times10$ at $U=10t,V=10t$ on the square lattice with Chern number $C=3$ and
  (b) the corresponding total Berry curvatures of the two $K=(0,0)$ ground states in the plane $(\theta^{x},\theta^{y})$.}
\end{figure}

Now we turn to analyze the relationship between the multicomponent FQH states at $\nu=N/(3N+1)$ and
the single-component FQH states at fillings $\widetilde{\nu}=1/(3C+1)$ of topological flat bands with Chern number $C=N$.
For topological bands with $C=N\geq1$, one can construct a Bloch-like basis in the $C$-component lowest Landau level,
and map an $N$-component FQH state to a corresponding state occurring on topological bands with $C=N$~\cite{YLWu2013}. Here, we consider the interacting Hamiltonian of hardcore bosons on the single layer square lattice with the lowest flat band hosting Chern number $C=N$, which can be obtained by twisting the $N$-layer checkerboard lattices as discussed in Ref.~\cite{Yang2012},
\begin{align}
&H^{[C=N]}=\sum_{\rr}\sum_{l=1}^{N}\left\{ t_{1}\left(b_{\rr+{\hat e_{x}}}^{[l+1]\dagger}+e^{i2l\phi}b_{\rr-{\hat e_{y}}}^{[l+1]\dagger}\right)b_{\rr}^{[l]}\right.\notag \\
 &\quad+t_{2}\left[e^{-i\left(2l-1\right)\phi}b_{\rr+{\hat e_{x}}+{\hat e_{y}}}^{[l]\dagger}+e^{i\left(2l-1\right)\phi}b_{\rr-{\hat e_{x}}-{\hat e_{y}}}^{[l]\dagger}\right. \notag \\
&\quad\left.\left.+e^{i\left(2l+1\right)\phi}b_{\rr+{\hat e_{x}}-{\hat e_{y}}}^{[l+2]\dagger}\right]b_{\rr}^{[l]}+\mathrm{H.c.}\right\}+ V_{int},\nonumber\\
&\quad\quad V_{int}=U\sum_{l\neq l'}\sum_{\rr}n_{\rr}^{l}n_{\rr}^{l'}+V\sum_{l,l'}\sum_{\langle\rr,\rr'\rangle}n_{\rr}^{l}n_{\rr}^{l'},\nonumber
\end{align}
where $n_{\rr}^{l}=b_{\rr}^{[l]\dagger}b_{\rr}^{[l]}$ the particle operator at sites $l=1,\cdots,N$.
Now each unit cell contains different $N$ layers. In Ref.~\cite{Zeng2018}, the bosonic FQH states at $\widetilde{\nu}=1/(N+1)$ are manifest up to $N=4$ under strong Hubbard interaction $U$, and here we consider the effect of nearest-neighbor interaction $V$ on bosonic FQH states at $\widetilde{\nu}=1/(3N+1)$. In the low-energy physics, when SU$(N)$ symmetric interactions among  different $N$ inequivalent sites within each unit cell are projected onto the lowest band with $C=N$, we can obtain the SU$(N)$ symmetric color-neutral projected Hamiltonian, and expect the emergence of the bosonic SU$(C=N)$ color-singlet FQH states at fillings $\widetilde{\nu}=1/(3N+1)$ for strong interactions $U,V\gg1$.

\begin{figure}[t]
  \includegraphics[height=1.4in,width=3.4in]{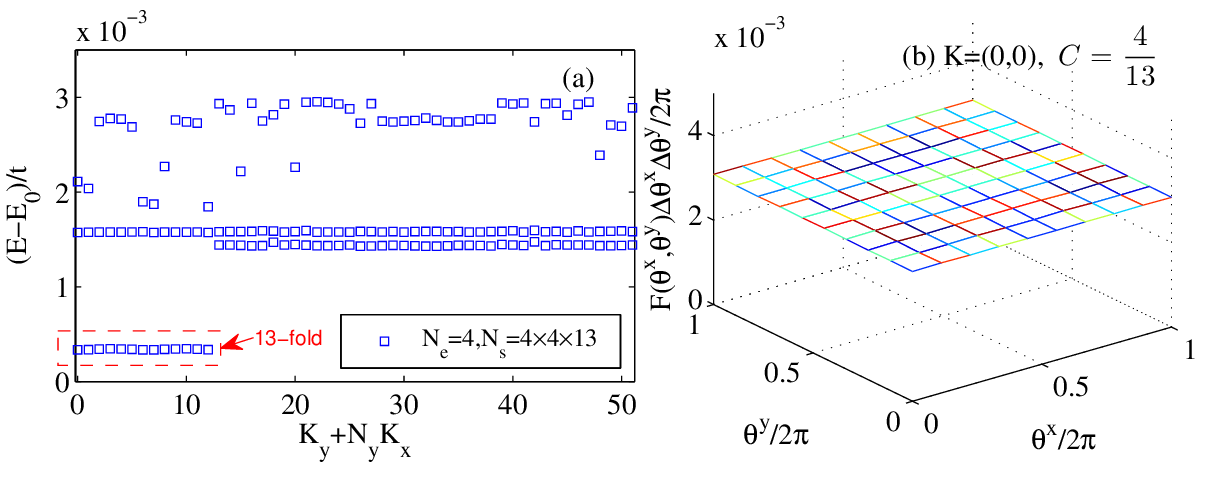}
  \caption{\label{energy4}(Color online) Numerical ED results for (a) the low-energy spectrum of hardcore bosons $\widetilde{\nu}=1/13,N_s=2\times N_e\times13$
  at $U=10t,V=10t$ on the square lattice with Chern number $C=4$ and (b) the corresponding Berry curvatures of the $K=(0,0)$ ground state in the plane $(\theta^{x},\theta^{y})$.}
\end{figure}

From Figs.~\ref{energy2}(a) to~\ref{energy4}(a), we plot the low-energy spectrum of strongly interacting hardcore bosons at filling
$\widetilde{\nu}=1/(3N+1)$ on the topological square lattice with Chern number $C=N$ ($N=2,3,4$). It is clear that, the ground states have $(3N+1)$-fold degeneracy.
In Figs.~\ref{energy2}(b) to~\ref{energy4}(b), using the twisted boundaries we numerically verify that the many-body Chern number equals to the Hall conductance $N\widetilde{\nu}=\nu=\sigma_H$.
Both the degeneracy and the Hall conductance match well with the predictions of the $\mathbf{K}$ matrix in Eq.~\ref{kmatrix}.
For larger system sizes, our DMRG calculation gives a nearly fractionally quantized charge pumping $\Delta N=C/(3C+1)=\sigma_H$
under the adiabatic insertion of one flux quantum, which demonstrates the robustness of these fractionalized phases.

Combined with the results of FQH states for $M=1,2$ in Refs.~\cite{Zeng2017,Zeng2018}, it is natural and
convincing to derive the general one-to-one correspondence between the $N$-component FQH states at $\nu=N/(MN+1)$ (odd $M$ for hardcore bosons and even $M$ for fermions)
on the topological lattice with unit Chern number, and the single-component FQH states at $\widetilde{\nu}=1/(MN+1)$ on the topological lattice with Chern number $N$.

\section{Drag Hall conductance and charge pumping}\label{drag}

\begin{figure}[t]
  \includegraphics[height=1.58in,width=3.4in]{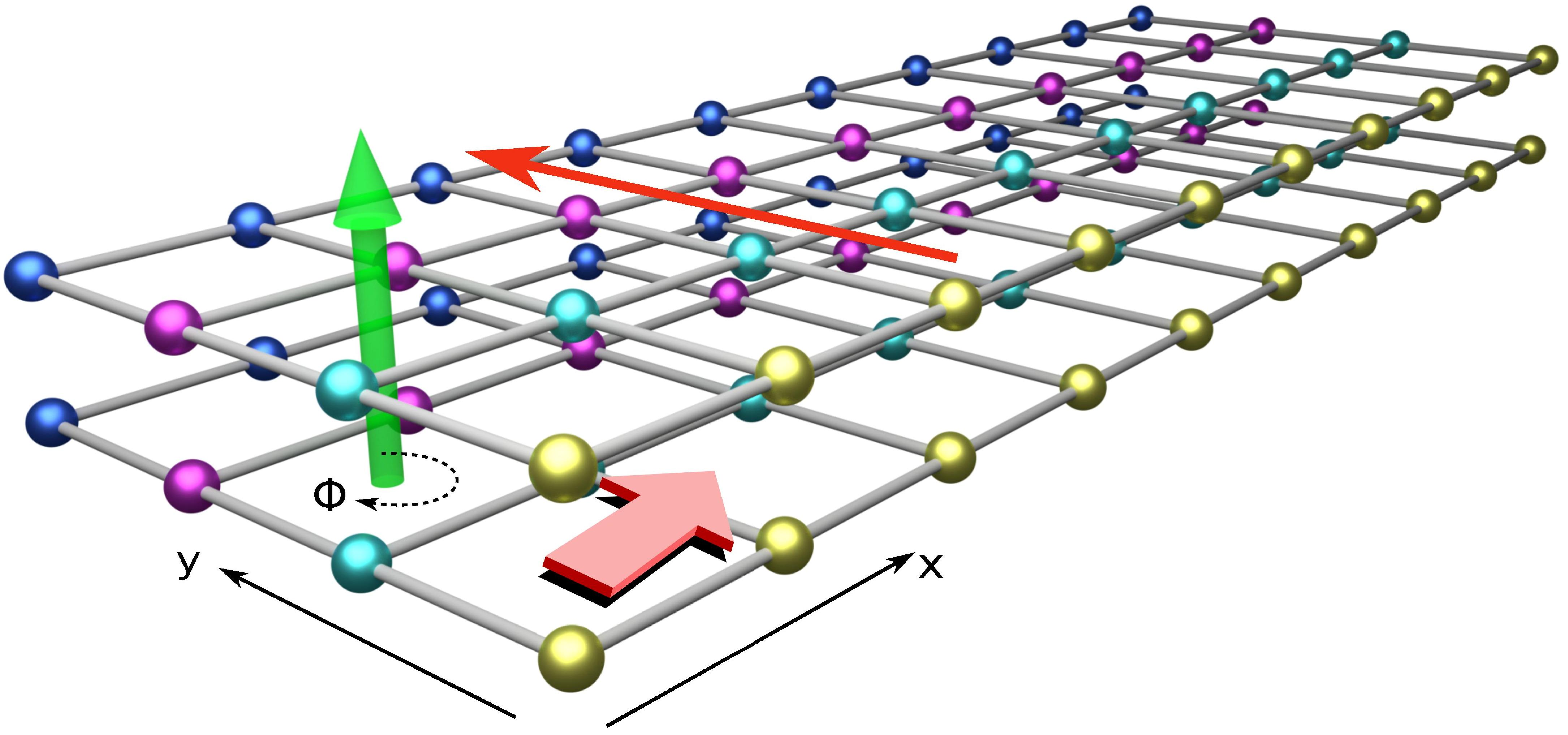}
  \includegraphics[height=1.52in,width=3.4in]{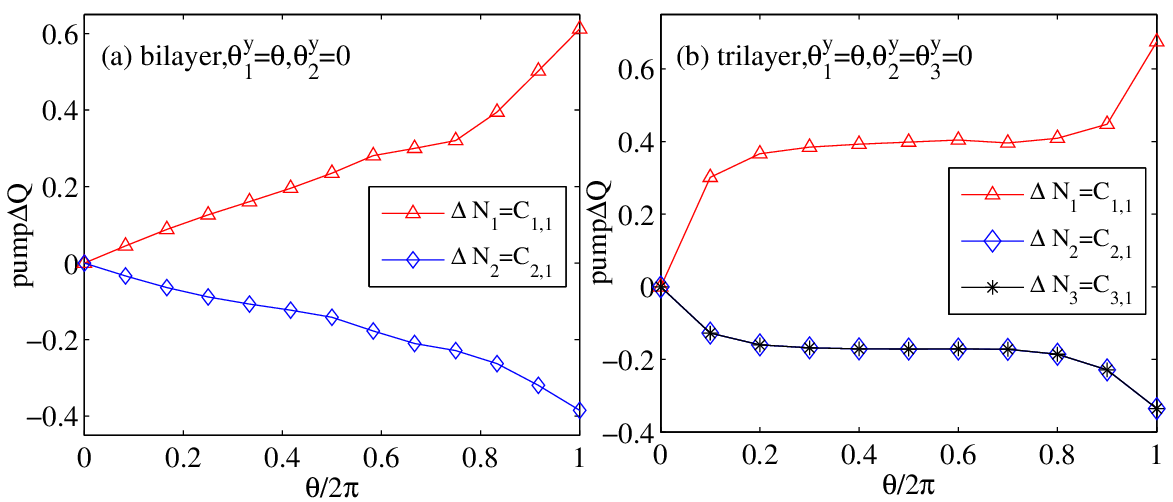}
  \caption{\label{pump}(Color online)
  (Top panel) Illustration of physical pictures of two-component system and the idea of drag charge pumping.
By applying an electric field (thick pink arrow) along the $x$ direction in one component (bottom layer),
it drives a charge pumping (thin red arrow) along the $y$ direction in the other component (top layer),
which generates the drag Hall conductance.
The charge transfer on the $N_y=3$ cylinder for (a) two-component bosons at $\nu=2/7,U=0,V=10t$ with inserting flux $\theta_{1}^y=\theta,\theta_{2}^y=0$ in a topological $\pi$-flux checkerboard lattice with cylinder length $L_x=N_x=28$; (b) three-component bosons at $\nu=3/10,U=0,V=10t$ with inserting flux $\theta_{1}^y=\theta,\theta_{2}^y=\theta_{3}^y=0$ in topological $\pi$-flux checkerboard lattice with cylinder length $L_x=N_x=30$. Here the calculation is performed using finite DMRG, and the maximal kept number of states is 2400.}
\end{figure}

As remarked above, the existence of off-diagonal elements $C_{\sigma,\sigma'}$ implies the quantized drag Hall responses in multicomponent systems; that is, when applying a driving force in one component, the Hall current will be observed in the other components as well.
To simulate this effect, we consider the topological charge pumping of the $\sigma$-th component in the $x$ direction
under the insertion of flux quantum $\theta_{\sigma'}^{y}$ of the $\sigma'$-th component in the $y$ direction as illustrated in Fig.~\ref{pump}.
With the help of DMRG, we can visualize such charge pumping in the ground state by continuously evolving the
ground state with the increasing of inserted flux~\cite{Gong2014},
akin to the experimental setup.
Technically, we partition the cylinder into two halves with equal lattice sites by a cut along $y$ direction.
%By threading flux field $\theta_{\sigma}^{y}=\theta,\theta_{\sigma'}^{y}=0$ from $\theta=0$ to $\theta=2\pi$,
%the expectation value of the particle number of the $\sigma$-th component on the left side equals to $N_{\sigma}^{L}(\theta)=tr[\widehat{\rho}_L(\theta)\widehat{N}_{\sigma}^{L}]$,
%where $N_{\sigma}^{L}$ is the particle number of spin-$\sigma$ in the left cylinder part, and $\widehat{\rho}_L$ is the reduced density matrix of the corresponding left half of the system.
The dynamical change of the particle number on the left side will be related to the net charge transfer across the bipartition entanglement cut; that is,
%The net charge transfer of the $\sigma'$-th component particle from the right side to the left side during each flux is
\begin{align}
  \Delta N_{\sigma'}=N_{\sigma'}^{L}(2\pi)-N_{\sigma'}^{L}(0)=C_{\sigma',\sigma},
\end{align}
where $N_{\sigma}^{L}$ is the particle number of the $\sigma$ component in the left cylinder part.

\begin{figure}[t]
  \includegraphics[height=1.52in,width=3.4in]{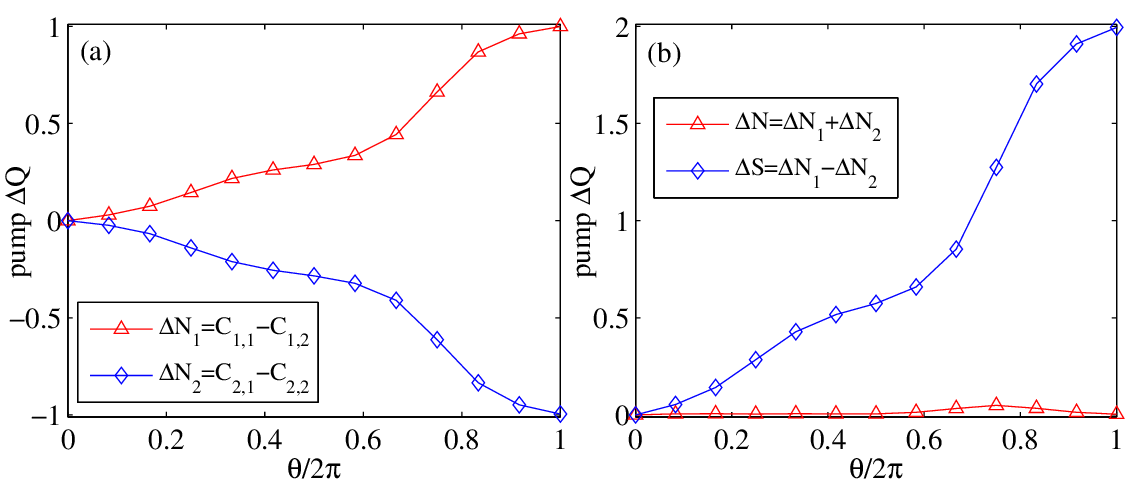}
  \caption{\label{spin}(Color online) The charge transfer on the $N_y=3$ cylinder for two-component bosons at $\nu=2/7,U=0,V=10t$ with inserting flux $\theta_{1}^y=-\theta_{2}^y=\theta$ in a topological $\pi$-flux checkerboard lattice with cylinder length $L_x=N_x=28$. (a) The charge pumping of each component system. (b) The total charge pumping and spin pumping. Here the calculation is performed using finite DMRG, and the maximal kept number of states is 2400.}
\end{figure}

As shown in Fig.~\ref{pump}(a), for two-component bosons at $\nu=2/7$, a fractional charge $\Delta N_1\simeq4/7=C_{1,1}$ is pumped in one component where the flux is inserted,
and a fractional charge $\Delta N_2\simeq-3/7=C_{2,1}$ is pumped in other component by threading one flux quantum $\theta_{1}^y=\theta,\theta_{2}^y=0$, demonstrating its Chern number matrix $\mathbf{C}=\frac{1}{7}\begin{pmatrix}
4 & -3\\
-3 & 4\\
\end{pmatrix}$, consistent with the analysis of ED study.
Similarly, for three-component bosons at $\nu=3/10$, by threading one flux quantum $\theta_{1}^y=\theta,\theta_{2}^y=\theta_{3}^y=0$,
a fractional charge $\Delta N_{1}\simeq0.7=C_{1,1}$ is pumped as the intra-species pump, and a fractional charge $\Delta N_{2}=\Delta N_{3}\simeq-0.3=C_{2,1}=C_{3,1}$ pumped in the other species,
as indicated in Fig.~\ref{pump}(b), demonstrating the Chern number matrix and its inverse $\mathbf{K}$ matrix formula in Eq.~\ref{kmatrix} for $M=3,N=3$:
\begin{align}
  \mathbf{C}=\frac{1}{10}\begin{pmatrix}
7 & -3 & -3\\
-3 & 7 & -3\\
-3 & -3 & 7\\
\end{pmatrix}, \mathbf{K}=\mathbf{C}^{-1}=\begin{pmatrix}
4 & 3 & 3\\
3 & 4 & 3\\
3 & 3 & 4\\
\end{pmatrix}.
\end{align}
Interestingly, for two-component bosons at $\nu=2/7$, one can also define the total charge pumping $\Delta N$ and spin pumping $\Delta S$ by
\begin{align}
  \Delta N &=\Delta N_1+\Delta N_2,\label{drag1}\\
  \Delta S &=\Delta N_1-\Delta N_2.\label{drag2}
\end{align}
Under the insertion of flux quantum $\theta_{1}^y=\theta_{2}^y=\theta$, we obtain the total charge pumping $\Delta N=\sum_{\sigma,\sigma'}C_{\sigma,\sigma'}=\nu$, namely, the charge Hall conductance, while the spin pumping $\Delta S=C_{1,1}+C_{1,2}-C_{2,1}-C_{2,2}=0$ during each cycle. In contrast, by threading one flux quantum $\theta_{1}^y=-\theta_{2}^y=\theta$, we obtain the charge pumpings $\Delta N_1=C_{1,1}-C_{1,2}$ and $\Delta N_2=C_{2,1}-C_{2,2}=-\Delta N_1$ in Fig.~\ref{spin}(a); that is, the two-component particles move exactly in the opposite directions. Hence the total charge pumping $\Delta N=\Delta N_1+\Delta N_2=0$ while the spin pumping is quantized to $\Delta S=\Delta N_1-\Delta N_2=2$, as shown in Fig.~\ref{spin}(b). Experimentally, one can apply spin-dependent driving forces (e.g. spin-dependent superlattice potentials~\cite{Schweizer2016} and valley-dependent electric driving fields~\cite{Spanton2017}), and obtain the drag Hall conductance from the counterflow of two-component particles from Eqs.~\ref{drag1} and~\ref{drag2}.

\section{Summary and Discussions}\label{summary}

In summary, by numerically exposing the topological Chern number matrix of the multicomponent systems, we
have demonstrated the topological characterization of the $N\times N$ $\mathbf{K}$ matrix of the bosonic SU$(N)$
FQH states at a partial filling $\nu=N/(3N+1)$ of the lowest Chern band with unit Chern number,
based on topological properties including the ground state degeneracy and fractional charge pumpings.
We also established the close relationship of such states to the single-component FQH states at fractional fillings $\widetilde{\nu}=1/(3N+1)$ of
the lowest Chern band with high Chern number $N>1$ on the topological lattice models.
In combination with $M=1,2$ obtained in Refs.~\cite{Zeng2017,Zeng2018}, our results reveal a large sequence of  SU$(N)$ FQH states at a partial filling $\nu=N/(MN+1)$
(odd $M$ for bosons and even $M$ for fermions).
The sequential fermionic FQH states with $M=4$ at filling $\nu=N/(4N+1)$, but not discussed here, are left for study in the near future.
As a final remark, we note that the drag Hall conductance has a topological nature and can
be probed by cold atom experiments soon.
The demonstration of such a kind of fractional quantized charge transfer reveals and characterizes
the internal structure of topology of multicomponent systems.

\begin{acknowledgements}
This work is supported by start-up funding from Westlake University.
D.N.S was supported by National Science Foundation, PREM DMR-1828019.
D.N.S also acknowledges travel support from Princeton MRSEC through the National Science Foundation, Grant No. DMR-1420541.
\end{acknowledgements}

\end{document}